\documentstyle[twoside,fleqn,espcrc2,epsf]{article}

\title{Hadron correlators with improved fermions}

\author{UKQCD Collaboration: P.W.Stephenson\address{Department of
Physics, University of Wales, Swansea, \\
        Singleton Park, Swansea, SA2 8PP, UK}\hfill SWAT/50\\
Talk given at Lattice '94, Bielefeld\hfill hep-lat/9411022}

\begin{document}

\begin{abstract}
We investigate point-to-point correlation functions for various
mesonic and baryonic channels using the ${\cal O}(a)$-improved Wilson
action due to Sheikholeslami and Wohlert.  We
consider propagators to both time slices 0 and 1.  We find that
discretisation effects are more pronounced than those reported with
unimproved Wilson fermions, but that the same procedure for removing
finite size effects is successful.  Extrapolating to the chiral limit,
we see the notable features predicted phenomenologically:  the ratio
of interacting to free correlators in the vector channel is roughly
constant to about 1 fm, while in the pseudoscalar channel the ratio
increases markedly due to the strong binding.
\end{abstract}

\maketitle

\section{Introduction}

Traditionally, much of lattice gauge theory has between concerned with
calculations of on-shell quantities such as masses and matrix
elements.  However, it has been emphasised by Shuryak~\cite{Shu93}\
that there is a lot of interesting physics to be had by looking at
off-shell quantities.  In the review cited, the fundamental objects of
interest are space-like correlations between quantum fields in the
vacuum:
\label{eqn:corrs}
\begin{equation}
K(r_1-r_2) \equiv \langle0\vert TJ(r_1)\bar J(r_2)\vert0\rangle,
\end{equation}
where $J(r)$ are currents corresponding to some suitable quantum
numbers:  in meson channels, $J(r)=\bar\psi(r)\Gamma\psi(r)$, while in
baryon channels,
$J(r)=\epsilon_{ijk}\psi(r)\Gamma_1\psi(r)\Gamma_2\psi(r)$, for some
appropriate Dirac structure $\Gamma$, $\Gamma_1$ and $\Gamma_2$.

These correlations decay very fast; at large volume the dominant
behaviour is exponential in the lightest bound-state mass.  It is
therefore convenient to normalise them against the same correlations
calculated for non-interacting quarks, in other words using a frozen
gauge configuration.  We therefore define
\label{eqn:R}
\begin{equation}
R(r) \equiv {K_{\rm interacting}(r)\over K_{\rm free}(r)}.
\end{equation}

The correlations defined in this fashion contain in principle a great
deal of information about all that is happening in the chosen channel.
In this way we can show up, for example, the striking differences
between the vector and pseudoscalar which are far from obvious if one
looks only at masses or decay constants, and hence shed more light on
the mechanisms underlying confinement and chiral symmetry breaking.

A detailed analysis of point-to-point correlations has been carried
out by Chu et al.~\cite{Chu93}; we refer the reader to that paper for
more detailed definitions, formulae and methods.  Much of our
methodology is inherited from theirs.  Here we describe calculations
using Sheikholeslami-Wohlert $O(a)$-improved (S-W) fermions~\cite{She85}.  In
addition, our lattice is slightly finer than that of
ref.~\cite{Chu93}, though our lattice size is correspondingly smaller.

We shall now describe our data and its analysis and comparisons with
phenomenological results, then proceed to conclusions.


\section{Analysis}

Our data sets are on $24^3\cdot48$ quenched lattices at inverse
coupling $\beta=6.2$.  The quark propagators, inverted using the S-W
action, are at three hopping parameters $\kappa=0.14144$, $0.14226$
and $0.14262$.  On this lattice, $\kappa_{\rm criticial}=0.14314$.  We
have currently 27 configurations at all three $\kappa$; for
consistency we restrict ourselves here to these 27.  The S-W action
adds a two-link part to the Wilson action and also requires a rotation
of the fermion fields.  This is known to improve the perturbative
scaling behaviour~\cite{Hea91}; however, there is no particular reason
to suppose the type of analysis we are performing will benefit from
this.

We then sum over all rotationally equivalent points on all
configurations (taking account of antisymmetry in our baryon
correlations), using a bootstrap over all data to generate the errors.
We take the ratio of these with free propagators generated using an
adaption of the method of Carpenter and Baillie~\cite{Car85}.
Initially we used only propagators from the origin to all sites at
$t=0$, i.e. on the same time slice as the origin.




We use the method of ref.~\cite{Bur94} to improve finite size effects
by removing images of the source at the origin.  This works well.




However, we have no equivalent quantitative way of dealing with
anisotropy.  Examination of the data shows that there are strong
effects up to approximately six lattice units; these appear to be
significantly larger than those seen by Chu et al.\ with the Wilson
action~\cite{Chu93}.  These authors point out that propagators to
points nearest the body-diagonals of the lattice include more paths
and should therefore be nearer the continuum values. In our case it
appears that even these points are suffering from significant
anisotropy for distances $r < 6$.

\begin{figure}[tb]
\epsfxsize = 3in
\epsfbox{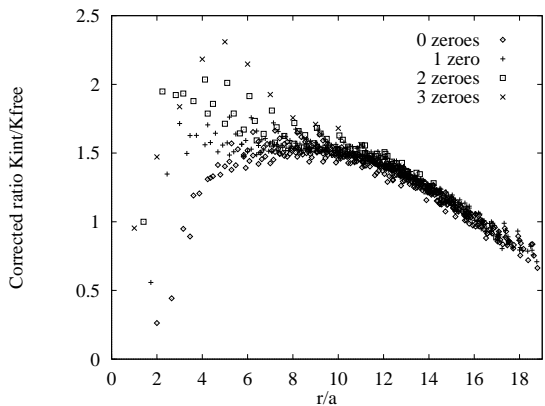}
\smallskip
\caption{Anisotropy effects in the vector channel}
\label{fig:anis}
\end{figure}
To shed further light on this, we calculate correlators to points on
the next time slice, $t=1$.  We have verified directly that the effect
of the differing lattice size in the time direction has negligible
impact on the correlations.  We find that these points lie
consistently lower than the $t=0$ points, so much so that it is highly
unlikely that the `more paths means lower anisotropy' ansatz can be
simply applied in our case.  On the other hand, the interpretation
`more paths means lower correlations' does seem to hold: in
figure~\ref{fig:anis} we show $R(r)$ from \ref{eqn:R}\ with the image
correction applied.  The different symbols denote co-ordinates with
different numbers of zeroes: i.e., $(x,y,z,t)$, $(x,y,z,0)$, etc.,
with $x$, $y$, $z$ and $t$ non-zero.  The expectation is that the
ratio remains roughly constant for small $r$ (note that this is for
quarks around the strange mass, for which the behaviour of the
anisotropy is cleanest), in other words that from $r=0$ to 8 the ratio
should be $\sim1.5$.  There is no easy way to make the data fit this.
Presumably this is an effect primarily of the rotation of the fermion
fields.  Indeed, the major additional problem appears to be that our
correlations plummet for some small values of $r$; apart from that the
anisotropy is similar to the Wilson case.

Matters are improved slightly by the choice of Chu et al.\ to
restrict points to those making an angle $\theta$ with the body
diagonal of the spatial lattice with $\cos(\theta)>0.9$.  In this case
the scatter of points is consistent with the statistical scatter from
about $r\sim6$, and our fits are performed from this distance on.

\subsection{Chiral extrapolations}

We follow the usual UKQCD convention in extrapolating our results to
zero quark mass, rather than the alternative of extrapolating to the
physical pion mass.  Our chiral extrapolations are limited by the fact
that currently we only have three quark masses.  We have tried various
forms of chiral extrapolation and have settled on a fit which linearly
extrapolates $\log(R(r))$.



\begin{figure}[thb]
\epsfxsize = 3in
\epsfbox{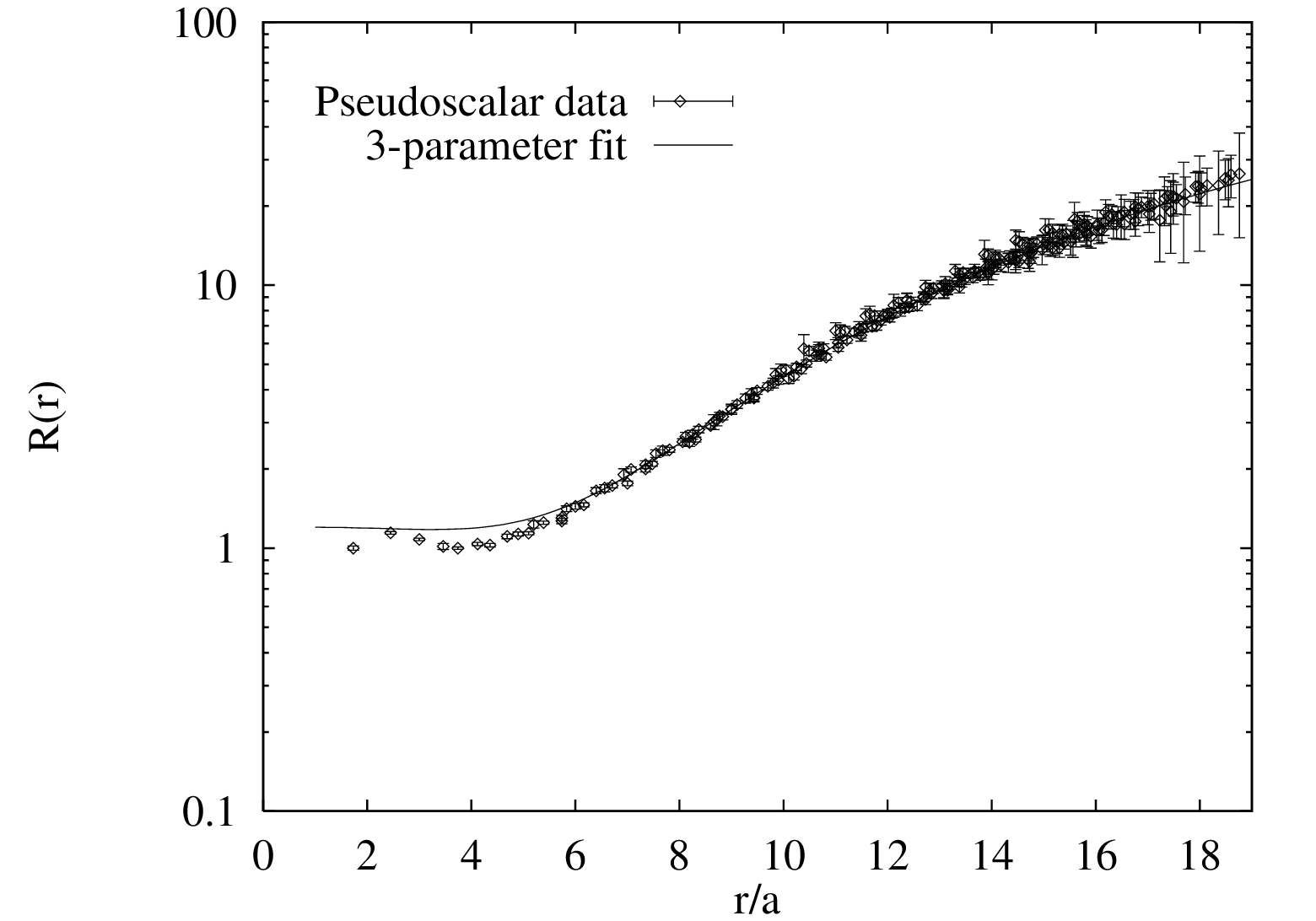}
\epsfxsize = 3in
\epsfbox{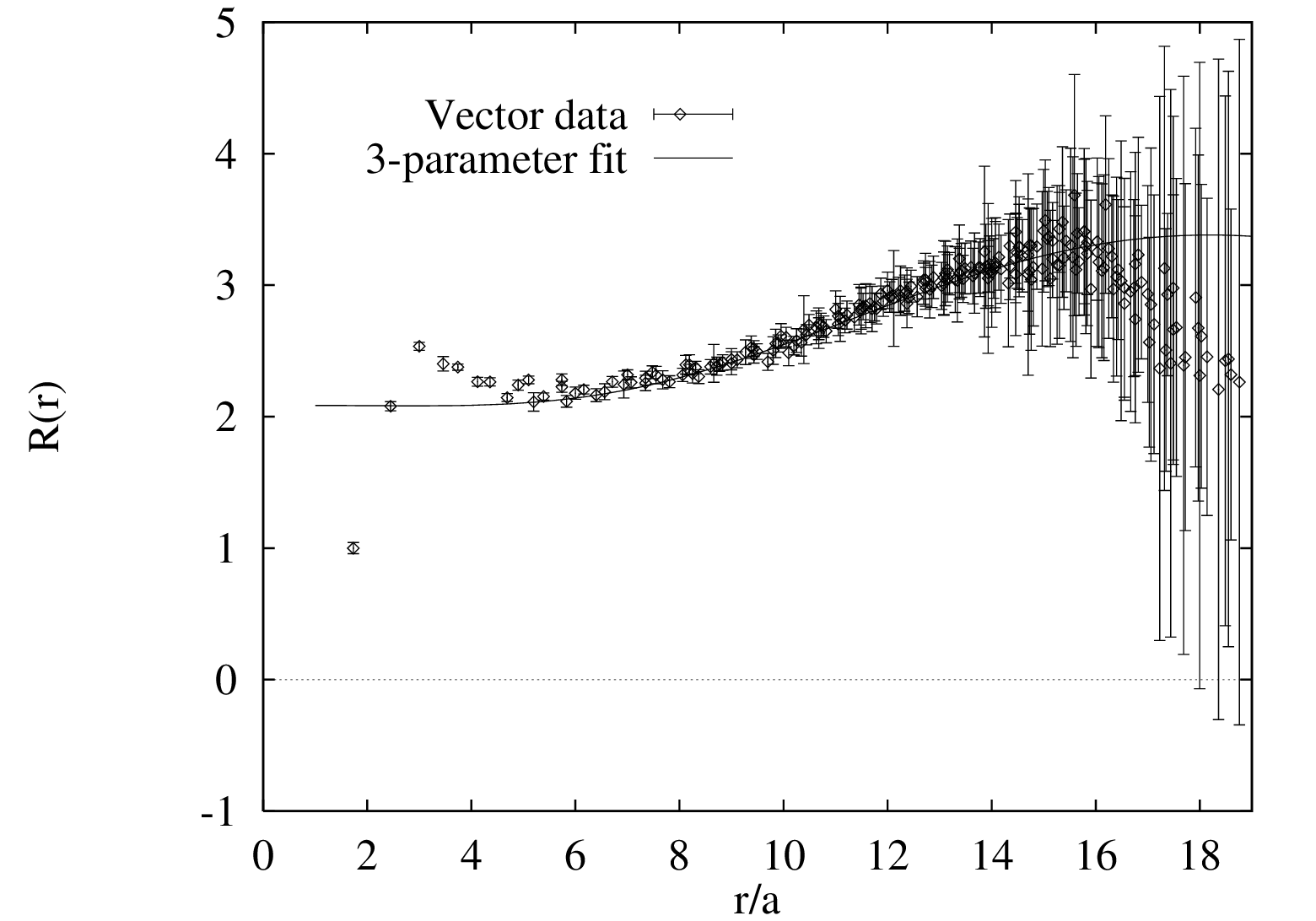}
\epsfxsize = 3in
\epsfbox{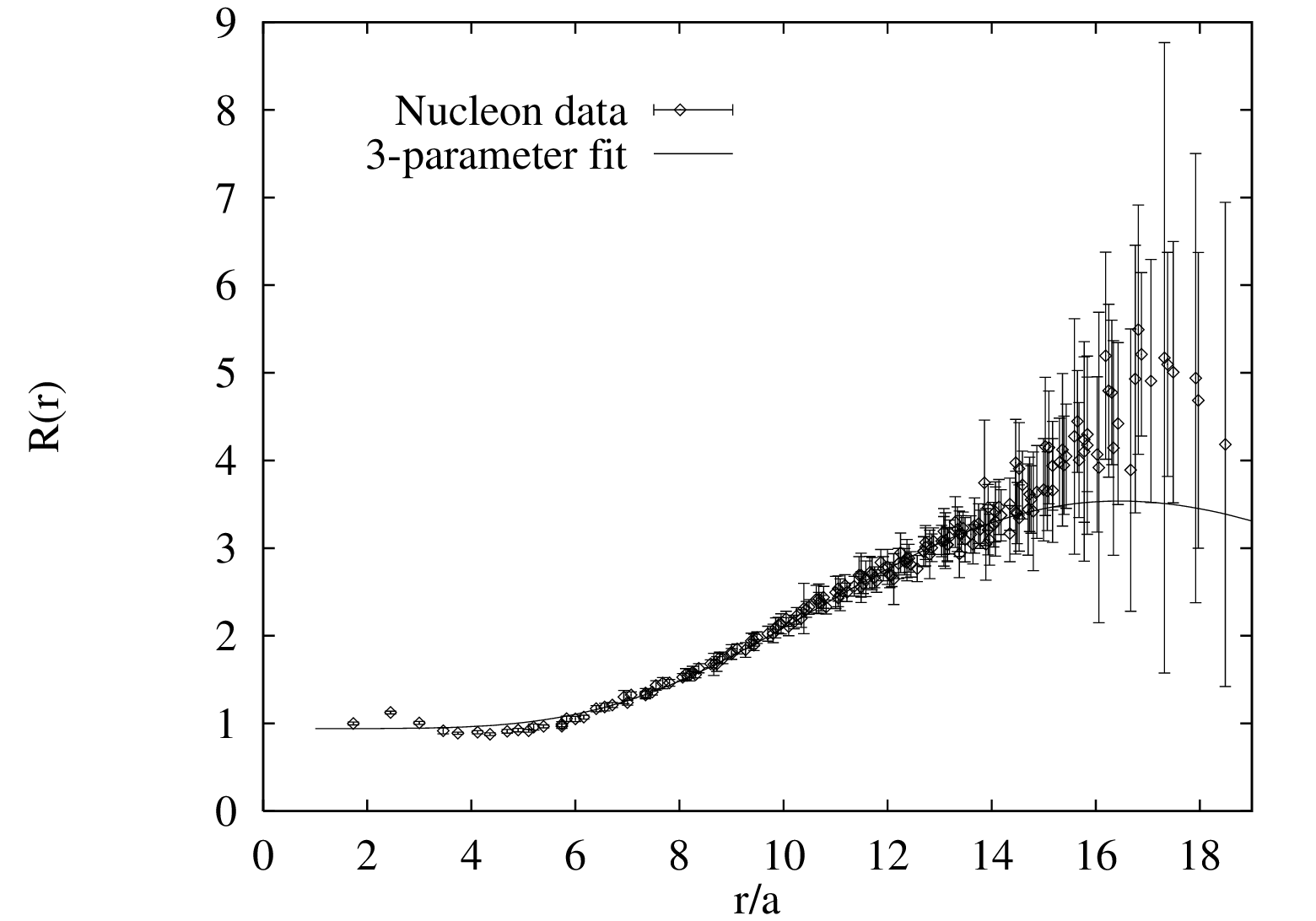}
\caption{Extrapolated data and fits for the pseudoscalar, vector and
nucleon.  The vertical scale is arbitrary:  anisotropy makes
normalisation difficult.}
\label{fig:fits}
\end{figure}

\subsection{Fits to phenomenological forms}

A reasonable phenomenological approximation to the data should be
found be assuming a smooth continuum background at large energy plus
some number of resonance poles.  In practice one pole is entirely
adequate and it is unlikely that our data could be persuaded to yield
a convincing second pole.  For the channels shown here this gives four
parameters to fit: an overall normalisation, the position ($m$) and
height ($\lambda$) of the delta function pole, and the threshhold for the
continuum.  Because we cannot fit for $r<0.6$ the estimation of the
continuum threshhold is generally poor and the fits shown just use a
reasonable fixed value.  Details of the forms are as in
ref.~\cite{Chu93}.  Errors are found by jackknife, completely
re-analysing the data with each set missing in turn.

We show the extrapolated data with the $\cos(\theta)>0.9$ cut off and
fits for the vector, nucleon and pseudoscalar in
figure~\ref{fig:fits}.  One fermi corresponds to about fourteen of our
lattice spacings.    We find a slightly low
rho mass ($620\pm90$ MeV) and a high nucleon mass ($1170\pm70$ MeV);
also the pseudoscalar is rather heavy ($300\pm20$ MeV), throwing some
doubt on the validity of extrapolating to zero quark mass for this
type of analysis.

\section{Conclusions}

We have examined point-to-point hadron correlation functions using an
$O(a)$ improved action.  We find more severe finite lattice spacing
effects than reported using Wilson fermions.  Nonetheless, the
physical features of the results are similar:  asymptotic freedom in
the vector channel persists to about 1 fm, but is violated quickly in
the pseudoscalar channel.


\begin{thebibliography}{99}
\bibitem{Shu93} E. Shuryak, Rev.\ Mod.\ Phys.\ 65 (1993) 1.
\bibitem{Chu93} M.-C. Chu, J.M. Grandy, S. Huang and J.W. Negele,
Phys.\ Rev.\ D48 (1993) 3340.
\bibitem{She85} B. Sheikholeslami and R. Wohlert, Nucl.\ Phys.\ B259
(195) 572.
\bibitem{Hea91} G. Heatlie, C.T. Sachrajda, G. Martinelli, C. Pittori
and G.C. Rossi, Nucl.\ Phys.\ B352 (1991) 266.
\bibitem{Car85} D.B. Carpenter and C.F. Baillie, Nucl.\ Phys.\ B260
(1985) 103.
\bibitem{Bur94} M. Burkardt, J.M. Grandy and J.W. Negele, `Calculation
and interpretation of hadron correlation functions in lattice QCD',
MIT preprint MIT-CTP-2109, hep-lat/9406009
\end{thebibliography}
\end{document}